\documentclass[conference]{IEEEtran}
\ifCLASSINFOpdf
\else
\fi
\usepackage{soul}
\usepackage{authblk}
\usepackage{bm}
\usepackage{xcolor}
\usepackage{stackengine}

\usepackage{cite}
\usepackage{amssymb}
\usepackage{mathtools}
\usepackage{mathrsfs}
\usepackage{eucal}
\usepackage{amsmath}
\usepackage{mathptmx}

\usepackage{graphicx}
\usepackage{pgfplots}
\usepackage{tikz}
\usetikzlibrary{patterns}
\usepackage{array}
\usepackage{subfigure}

\graphicspath{ {./images/} }
\usepackage[utf8]{inputenc}
\usepackage{comment}
\usepackage[utf8]{inputenc}
\usepackage{multirow}
\usepackage{url}
\usepackage{multicol}
\usepackage{xcolor}
\usepackage[linesnumbered,ruled,vlined]{algorithm2e}
\usepackage{algorithmic}

\SetCommentSty{mycommfont}
\SetKwInput{KwInput}{Input}
\SetKwInput{KwInitialize}{Initialize}
\SetKwInput{KwOutput}{Output} 

\begin{document}
\title{Dealing with CSI Compression to Reduce Losses and Overhead: An Artificial Intelligence Approach}
\author[1, 2]{Muhammad Karam Shehzad}
\author[1]{Luca Rose}
\author[2]{Mohamad Assaad}
\affil[1]{Nokia Bell Labs\\France.}
\affil[2]{Laboratoire des Signaux et Systemes, CentraleSupelec, Gif-sur-Yvette, France.}
\affil[ ]{Emails: muhammad.shehzad@nokia.com, luca.rose@nokia-bell-labs.com, mohamad.assaad@centralesupelec.fr}
\maketitle

\begin{abstract}
Motivated by the issue of inaccurate channel state information (CSI) at the base station (BS), which is commonly due to feedback/processing delays and compression problems, in this paper, we introduce a scalable idea of adopting artificial intelligence (AI) aided CSI acquisition. The proposed scheme enhances the CSI compression, which is done at the mobile terminal (MT), along with accurate recovery of estimated CSI at the BS. Simulation-based results corroborate the validity of the proposed scheme. Numerically, nearly 100\% recovery of the estimated CSI is observed with relatively lower overhead than the benchmark scheme. The proposed idea can bring potential benefits in the wireless communication environment, e.g., ultra-reliable and low latency communication (URLLC), where imperfect CSI and overhead is intolerable.      
\end{abstract}
\IEEEpeerreviewmaketitle
\section{Introduction} \label{sec1}

In wireless communication, in particular multiple-input multiple-output (MIMO)-based wireless networks, channel state information (CSI) is indispensable to provide high data rate. Accurate CSI at the transmitter allows efficient precoding, optimal transmit power allocation, modulation and coding scheme selection, etc. However, the acquisition of CSI is a crucial job. For example, in the fifth-generation (5G) environment, several communication technologies, e.g., unmanned aerial vehicle (in high mobility), and millimeter wave (with shorter wavelength), are prone to excessive channel fading \cite{karamUAV, rida}.
\par Albeit time-division-duplex (TDD) systems could exploit reciprocity to acquire CSI, in most of current schemes, in both frequency-division-duplex (FDD) and TDD systems \cite{FDD,TDD}, base stations (BSs) transmit reference symbols (RS) to the mobile terminals (MTs) to estimate the channel, and the MTs feedback the CSI to the BS. To reduce over-the-air (OTA)-overhead, the CSI feedback is largely compressed. At present, the Third Generation Partnership Project (3GPP) considers two strategies, i.e., type-I CSI and type-II CSI for CSI reporting \cite{3GPP1, 3GPP2}. Nevertheless, the downside is that both strategies involve strong compression of the estimated CSI with consequent deterioration of CSI accuracy. Lastly, it is important to remark that the necessity of feedback grows in massive-MIMO (mMIMO), which makes it challenging to design a mMIMO system having a fewer number of feedback \cite{lim_fb}.     
\par Interestingly, artificial intelligence (AI) has paved the way in many applications of wireless communication, and it has emerged as a paradigm shift in future needs \cite{rose, jakob}. Therefore, owing to the above issues, we utilize AI to overcome the losses, which occur due to CSI compression. In addition, we try to minimize OTA-overhead by means of a channel predictor functionality. To this end, in particular, we consider twin-channel predictors, which are synchronized, at both ends of the communication system, and feedback is evaluated on the basis of prediction at MT. Besides, different from most of the existing work, the training of channel predictor is based on quantized version of estimated CSI rather than actual CSI. 
\par The remainder of the paper is organized as follows. In Section\,\ref{systemmodel}, system model is discussed. Conventional scheme is summarized in Section\,\ref{benchmark_scheme}. AI-enabled proposed scheme is explained in Section\,\ref{proposed}. The results and their analysis is presented in Section\,\ref{results}. Finally, conclusion is made in Section\,\ref{conclusion}.  
\par $Notations:$ In this paper, $[\cdot]^{T}$ indicates the matrix transpose. Additionally, matrices are represented by boldface upper-case, vectors with boldface lower-case, and scalars with normal lower-case. Furthermore, $\widehat{\mathbf{H}}$, $\widetilde{\mathbf{H}}$, and $\mathbf{H}$ represent the estimated, predicted and actual channel, respectively. 
\section{System Model}\label{systemmodel}
\subsection{Communication Scenario}
Consider a single-cell downlink point-to-point MIMO system, having $N_{t}$ transmit and $N_{r}$ receive antennas, respectively. The MT receives dedicated RSs, transmitted by the BS, to estimate the channel. Once the channel is estimated, the CSI is compressed to reduce OTA-overhead, and thus fed back to the BS. Further, let us assume that both network entities are equipped with a channel predictor function. Without loss of generality, MIMO system can be modelled as
\begin{equation}\label{MIMO_model}
    \mathbf{y}(t)=\mathbf{H}(t)\cdot\mathbf{s}(t)+ \bm{\gamma}(t)\:
\end{equation}
where $\mathbf{y}(t)=\big[y_{1}(t), y_{2}(t), ..., y_{N_{r}}(t)\big]^{T}$ represents the received signal at time $t$, which has a dimension $N_{r}\times 1$. In addition, $\mathbf{s}(t)=\big[s_{1}(t), s_{2}(t), ..., s_{N_{t}}(t)\big]^{T}$ and $\bm{\gamma}(t)$ denote the vectors of transmitted symbols and additive white Gaussian noise (AWGN), respectively. Moreover, $\mathbf{H}(t)=\big[h_{n_{r}n_{t}}(t)\big]_{N_{r}\times N_{t}}$ depicts the channel matrix, having dimension $N_{r}\times N_{t}$, and $h_{n_{r}n_{t}}\in \mathbb{C}^{1\times1}$ is the complex-valued flat-fading channel gain between $n_{t}$ transmit and $n_{r}$ receive antennas, where $1\leq n_{t}\leq N_{t}$ and $1\leq n_{r}\leq N_{r}$.
\subsection{Channel Model}
The time varying mobile radio channel can be modeled using an auto-regressive (AR) process \cite{autoregressive}, which is capable of generating future channel values by combining past realizations and AR coefficients. For the design of discrete-time simulation of the model, the auto-correlation function (ACF) is given by
\begin{equation}\label{auto}
R[n]= J_{0}(2\pi f_{m}|n|)\: 
\end{equation}
where $J_0(\cdot)$ represents the zeroth-order Bessel function. Additionally, $f_m=f_d\cdot T_s$ depicts the maximum Doppler frequency (in Hertz) normalized by the sampling rate, $1/T_s$. Also, $f_d$ is the simple maximum Doppler frequency (in Hertz), which can be written as $f_d= \varsigma/\lambda$, where $\varsigma$ is the speed of the mobile device and $\lambda$ is the carrier wavelength.
Without loss of generality, a $u^{th}$ order complex AR process, denoted by AR($u$), can be generated by
\begin{equation}
    {\zeta}[n]=\sum_{i=1}^{{u}} \varphi_i\cdot \zeta[n-i]+ \omega(n)\:
\end{equation}
where $\varphi_{i}$ represents a coefficient of AR model and $\omega(n)$ is the zero-mean complex AWGN having variance $\sigma_{u}^2$. 
Finally, AR coefficients, $[\varphi_1, \varphi_2, ..., \varphi_u]$, can be obtained by solving the set of $u$ Yule-Walker equations as
\begin{equation}
\mathbf{R}\bm{c}= \mathbf{r}\:
\end{equation}
where
\begin{equation}
    \mathbf{r}=\big[R[1], R[2], ..., R[u]\big]^{T}\:
\end{equation}
\begin{equation}
    \bm{c}=\big[\varphi_1, \varphi_2, ..., \varphi_u\big]^{T}\:
\end{equation}

\begin{gather}
\bm{R}=
  \begin{bmatrix}
    R[0]&R[-1]&\cdots & R[-u+1]\\
    R[1]&R[0]&\cdots & R[-u+2]\\
    \vdots&\vdots&\ddots&\vdots\\
    R[u-1]&R[u-2]&\cdots & R[0]
  \end{bmatrix}\:
  \end{gather}
  and
  \begin{equation}
      \sigma_u^{2}=R[0]+\sum\limits_{i=1}^{u} \varphi_i\cdot R[i]\:.
  \end{equation}
Thus, having AR coefficients, channel, at time $t$, can be obtained as
\begin{equation}
    h(t)=\sum\limits_{i=1}^{u} \varphi_i\cdot h(t-i)+\omega(t)\:.
\end{equation}
Similarly, for a MIMO channel, the above equation can be extended as
\begin{equation}
    \mathbf{H}(t)=\sum\limits_{i=1}^{u} \mathbf{C}_i\odot  \mathbf{H}(t-i)+\mathbf{\omega}(t)\:
\end{equation}
where $\odot$ represents the element-wise multiplication, i.e., Hadamard product, of two matrices. Additionally, $\mathbf{C}_i$ is written as
\begin{gather}
\mathbf{C}_{i}=
  \begin{bmatrix}
    \varphi_{i}^{1,1}&\varphi_{i}^{1,2}&\cdots & \varphi_{i}^{1,N_t}\\
    \varphi_{i}^{2,1}&\varphi_{i}^{2,2}&\cdots & \varphi_{i}^{2,N_t}\\
    \vdots&\vdots&\ddots&\vdots\\
    \varphi_{i}^{N_r,1}&\varphi_{i}^{N_r,2}&\cdots & \varphi_{i}^{N_r,N_t}
  \end{bmatrix}\:
  \end{gather}
which is the AR coefficient matrix having entry $\varphi_{i}^{n_r,n_t}$, which represents the $i^{th}$ coefficient of the AR model, for $n_r$ and $n_t$ receive and transmit antennas, respectively.
\par In the following section, we briefly summarize the conventional channel estimation scheme. Later on, we discuss the proposed scheme along with the potential benefits.  
\section{Conventional Scheme}\label{benchmark_scheme}
In the conventional scheme, the BS transmits dedicated RS, at time $t$, to the MT to get the estimate of the channel. Then, the MT estimates\footnote{e.g., in our work, we assume KF-based channel estimation \cite{KF2}.} the channel, denoted by $\widehat{\mathbf{H}}_{MT}$, using the received RS and transmits the feedback to the BS as
\begin{equation}
    \widehat{\mathbf{H}}^{q}= Q_f(\widehat{\mathbf{H}}_{MT})\:
    \label{conventional_scheme}
\end{equation}
where $Q_f(\cdot)$ denotes the quantization\footnote{Here, quantizer-function is a standard element-wise quantization.} function. Nonetheless, such quantized channel, $ \widehat{\mathbf{H}}^{q}$, can reduce the performance of the estimated channel and can also result in higher OTA-overhead. To overcome these problems, below, we address the proposed scheme. 
\section{Proposed Scheme}\label{proposed}
\begin{figure*} [ht]
\begin{center}
  \includegraphics[width=18cm,height=8cm]{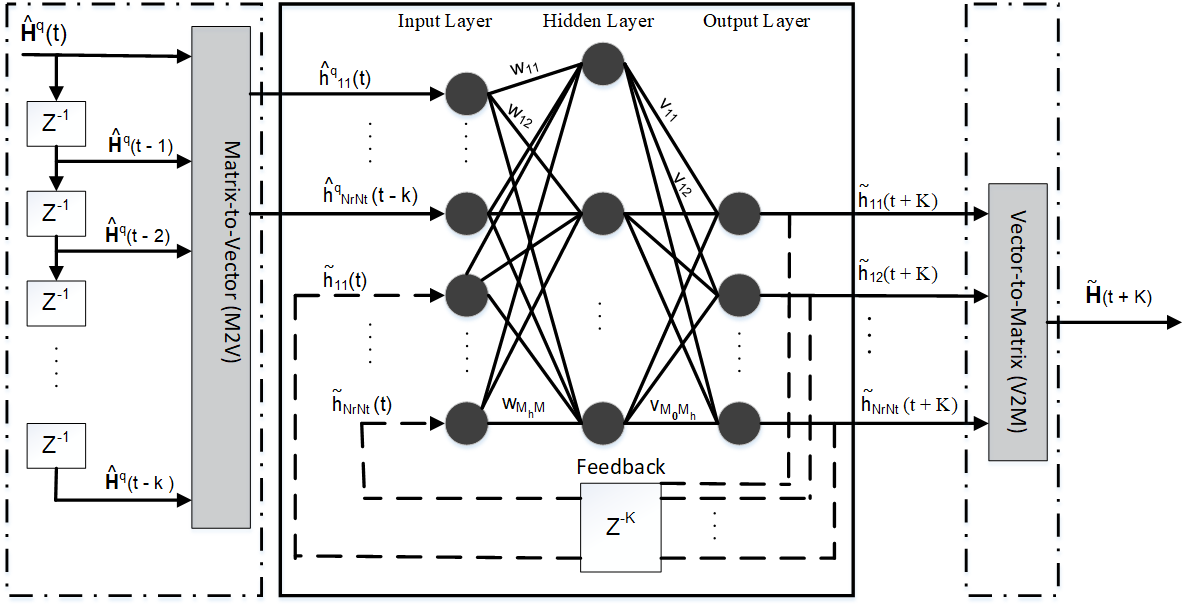}
  \caption{{Pictorial representation of RNN-based multi-step MIMO channel predictor.}}\label{RNN}      
\end{center}
\end{figure*} 
The proposed scheme considers the use of twin-channel predictors at both ends of the communication system, i.e., BS and MT. The key idea is to evaluate the feedback based on the prediction at the MT; thereby, reducing the number of feedbacks depending on the predicted channel. Moreover, reducing the necessary overhead to feedback the estimated channel from MT. Therefore, in the proposed scheme, firstly, we assume that through the received data, MT estimates the channel, $\widehat{\mathbf{H}}_{MT}$, using conventional scheme. Later on, the estimated channel is quantized and fed back to the BS. Consequently, the quantized estimated channel, denoted by $\widehat{\mathbf{H}}^{q}$, is available at both ends of the communication system. Furthermore, the proposed scheme consists of three phases: prediction phase, reporting phase, and the recovery phase (at BS). In the following subsections, we provide the details of each phase.
\subsection{Prediction Phase: AI-Enabled Channel Prediction}
During the prediction phase, both the network entities adopt AI to predict the next channel realization based on the previous channel realization. Importantly, both the channel predictors use $\widehat{\mathbf{H}}^{q}$ for training. In the following, we describe the AI-enabled channel predictor adopted in our work.
\par Within the domain of AI, recurrent neural network (RNN) is a type of AI algorithm that has the potential of predicting the time-series data \cite{RNN_paper}. Motivated by this capability, this work adopts a RNN-based channel predictor. The goal of the RNN-based predictor is to obtain $K$-step ahead prediction of the channel, denoted by $\widetilde{\mathbf{H}}[t+K]$, which is as close as possible to $\widehat{\mathbf{H}}^{q}[t+K]$. A typical RNN architecture to predict multi-step MIMO channel is drawn in Fig.\,\ref{RNN}. Specifically, Fig.\,\ref{RNN} depicts a typical multi-input multi-output RNN, which consists of three layers: an input layer, a hidden layer, and an output layer. In terms of AI, this architecture is well known as a single hidden layer RNN, or a two-layer RNN (input layer is generally excluded). Further, input layer has $M$ input neurons, which consist of external input and the feedback from the output. Similarly, the hidden layer has $M_{h}$ neurons, and output layer is composed of $M_{o}$ neurons.
\par At time $t$, the corresponding quantized channel, $\widehat{\mathbf{H}}^{q}(t)$, along with $k$-step delays (as shown with dotted-dashed box in left part of Fig.\,\ref{RNN}), i.e., $\widehat{\mathbf{H}}^{q}(t-1), \widehat{\mathbf{H}}^{q}(t-2), ..., \widehat{\mathbf{H}}^{q}(t-k)$, is fed as external input to the RNN. Nonetheless, in the context of AI, such input data should be pre-processed, i.e., unrolled into a vector, to feed into the RNN architecture. Therefore, we have drawn a pre-processing block, represented within right side of dotted-dashed box, which unrolls the $\widehat{\mathbf{H}}^{q}$ into vector as   
\begin{equation}
    \widehat{\mathbf{h}}^q= \text{M2V}(\widehat{\mathbf{H}}^{q})= [\widehat{h}_{11}^q\quad \widehat{h}_{12}^q \quad...\quad \widehat{h}_{N_rN_t}^q]\:
\end{equation}  
Here, for convenience, we use ${\mathbf{h}}$ as a vector. Alongside external input at time $t$, feedback or a recurrent component, represented by
\begin{equation}
    \widetilde{\mathbf{h}}(t)= [\widetilde{h}_{11}(t)\quad \widetilde{h}_{12}(t) \quad ... \quad \widetilde{h}_{N_rN_t}(t)]
\end{equation}
is fed as an internal input to the RNN. Therefore, combined input vector to the RNN at time $t$ is written as
\begin{equation}
\mathbf{x}(t)=[\widehat{\mathbf{h}}^q(t), \widehat{\mathbf{h}}^q(t-1), ..., \widehat{\mathbf{h}}^q(t-k), \widetilde{\mathbf{h}}(t)] \:.   
\end{equation}
Finally, the output of RNN is simply $K$-steps ahead prediction, denoted by $\widetilde{\mathbf{h}}(t+K)$, which can be transformed into the form of MIMO predicted channel, i.e., $\widetilde{\mathbf{H}}(t+K)$, using a post-processing block (denoted by dotted-dashed box on the extreme right side of Fig.\,\ref{RNN}). Thus, the predicted channel can be expressed as
\begin{gather}
\widetilde{\mathbf{H}}(t+K) =
  \begin{bmatrix}
    \widetilde{h}_{11}(t+K)&\cdots & \widetilde{h}_{1N_t}(t+K)\\
    \widetilde{h}_{21}(t+K)&\cdots & \widetilde{h}_{2N_t}(t+K)\\
    \vdots&\ddots&\vdots\\
    \widetilde{h}_{N_r1}(t+K)&\cdots &\widetilde{h}_{N_rN_t}(t+K)
   \end{bmatrix}\:.
\end{gather}
\par Importantly, the prediction behavior of a RNN is fully dependent on its weight values and the activation function. As depicted in Fig.\,\ref{RNN}, we denote the weights for $j^{th}$ hidden neuron with $m^{th}$ predecessor neuron as $w_{jm}$, whereas $v_{oj}$ represents the weight for $o^{th}$ output neuron, where $1 \leq m\leq M$, $1 \leq j\leq M_{h}$, and $1\leq o\leq M_{o}$. Moreover, generally, hyperbolic tangent ($tanh$) function is used as an activation function in the hidden layers, which is defined as
\begin{equation}
    tanh(z)=\frac{e^{z}-e^{-z}}{e^{z}+e^{-z}}\:.
\end{equation}
Therefore, the output activation of $j^{th}$ hidden neuron at time $t$ can be written as
\begin{equation}
    a_{j}(t)=tanh(\mathbf{w}_j\cdot\mathbf{x}(t))\:
    \label{activation}
\end{equation}
where $\mathbf{w}_j= [w_{j1}, ...,w_{jM}]$. Similarly, having the activation of predecessor layer, i.e., $a_{j}(t)$, the output for the $o^{th}$ neuron, which depicts the $K$-step ahead prediction for the particular communication channel, is given as
\begin{equation}
    y_o(t+K)=\sum_{j=1}^{M_{h}}v_{oj}\cdot a_j(t)\:
    \label{output}
\end{equation}
where in the context of our work, $o=n_{t}+(n_r-1)N_{t}$.
Finally, by substituting \eqref{activation} into \eqref{output}, $K$-step ahead prediction at a particular neuron, which depicts prediction for the channel, $\widetilde{h}_{n_rn_t}(t+K)$, can be expressed as
\begin{equation}
    \widetilde{h}_{o}(t+K)=\sum_{j=1}^{M_{h}} v_{(n_{t}+(n_r-1)N_{t})j}\cdot tanh(\mathbf{w}_j\cdot\mathbf{x}(t))\:.
\end{equation}
\par In order to predict future channel realizations, the RNN needs to be trained. Therefore, once the parameters, i.e., the number of neurons and layers, of the RNN have been chosen, the training process can begin by providing both training and validation (optional) data along with labels. At each training iteration, RNN evaluates the cost function and back propagates the error to update the weights. This process is repeated until the convergence condition is reached, i.e., the cost function has minimized. Once the RNN has been trained, it can be used to predict the channel. To get a deeper understanding of RNN, interested readers are recommended to read, e.g., \cite{RNN_paper, MohamadRNN}.
\par In the following, we discuss the channel reporting strategy, which is followed by the MT.
\subsection{Reporting Phase}\label{reporting}
Once the RNN training phase is completed, the predictors at both ends of the communication, generate the same\footnote{Here, it is, however, important to note that the predictor on both ends of communication system is same. Additionally, it is using the same data for training RNN at BS and MT; therefore, the prediction would be identical on both sides, i.e., $\widetilde{\mathbf{H}}_{MT}=\widetilde{\mathbf{H}}_{BS}$, where $\widetilde{\mathbf{H}}_{MT}$ and $\widetilde{\mathbf{H}}_{BS}$ denote the predicted channel at MT and BS, respectively.} channel, for instance, at time instant $t-1$. On the other hand, at time $t$, MT receives a dedicated RS to estimate the channel, $\widehat{\mathbf{H}}_{MT}(t)$. After the channel is estimated at MT, it computes the difference between the predicted channel, i.e., $\widetilde{\mathbf{H}}_{MT}(t-1)$, and the estimated channel as
\begin{equation}
    \overline{\mathbf{H}}(t)=\widetilde{\mathbf{H}}_{MT}(t-1)- \widehat{\mathbf{H}}_{MT}(t)\:.
\end{equation}
In the next step, the update obtained in the above equation, is quantized as
\begin{equation}
    \overline{\mathbf{H}}^{q}(t)=Q^{p}_{f}(\widetilde{\mathbf{H}}_{MT}(t-1)- \widehat{\mathbf{H}}_{MT}(t))\:
    \label{q_update}
\end{equation}
where $Q^{p}_{f}(\cdot)$ represents a quantization function. Later on, the quantized-update, $\overline{\mathbf{H}}^{q}(t)$, is reported to the BS. 
\subsection{Estimation at BS}
In the final phase, by the time BS gets the feedback, $\overline{\mathbf{H}}^{q}(t)$, from the MT, it has also predicted the channel, denoted by $\widetilde{\mathbf{H}}_{BS}(t-1)$. Therefore, BS estimates the channel, at time $t$, as
\begin{equation}
    \widehat{\mathbf{H}}_{BS}(t)=\widetilde{\mathbf{H}}_{BS}(t-1)-\overline{\mathbf{H}}^{q}(t)\:.
    \label{est_BS}
\end{equation}
By substituting \eqref{q_update} into \eqref{est_BS}, the above equation can be written as
\begin{equation}
    \widehat{\mathbf{H}}_{BS}(t)=\widetilde{\mathbf{H}}_{BS}(t-1)-Q^{p}_{f}(\widetilde{\mathbf{H}}_{MT}(t-1)- \widehat{\mathbf{H}}_{MT}(t))\:.
    \label{final_est_BS}
\end{equation}
\par The major benefits of using the above approach are as follows. Firstly, if the predicted, $\widetilde{\mathbf{H}}_{MT}(t-1)$, and the estimated channel, $\widehat{\mathbf{H}}_{MT}(t)$, at the MT, are the same, then there is no need to feedback anything, and in such case $\widehat{\mathbf{H}}_{BS}(t)=\widetilde{\mathbf{H}}_{BS}(t-1)$; thus, feedback-related overhead is eliminated. For example, in an extreme ideal scenario, this can happen in the environment where MT is immobile (e.g., watching football match in a stadium, in the global pandemic (coronavirus), user sitting at home most of the time, etc.), or when variation in the channel is very low (e.g., MT is walking in the street), etc. Nonetheless, in static scenarios, only little variations may occur; thus, requiring only marginal updates. Hence, in such kind of scenarios, the overhead can be largely reduced. Secondly, if there is a difference between $\widetilde{\mathbf{H}}_{MT}(t-1)$ and $\widehat{\mathbf{H}}_{MT}(t)$, then the quantization in \eqref{q_update} will introduce less noise as compared to \eqref{conventional_scheme} (i.e., conventional scheme). Thereby, more precise estimated channel can be reported to BS. Lastly, for the proposed scheme, we conclude a few remarks, which are given below.
\begin{enumerate}
    \item The key point of proposed scheme is to either reduce the amount of quantization bits necessary for the feedback, i.e., \eqref{q_update} will require less bits than \eqref{conventional_scheme} for similar performance, or to increase the performance with the same amount of bits. 
    \item If the prediction was perfect at MT, then no feedback would be necessary; thereby, bringing the amount of necessary bits to 0.
    \item If infinite amount of bits are considered for the feedback, it is possible to define $Q_f(\beta)=Q^{p}_f(\beta)=\beta$, where $\beta$ is the data to be quantized. In other words, this would imply that  $\widehat{\mathbf{H}}_{BS}(t)=\widehat{\mathbf{H}}_{MT}(t)$, i.e., the proposed scheme gives no advantage. Correspondingly, the largest amount of gain is acquired with low resolutions feedback. This, in particular, is relevant as 3GPP CSI acquisition schemes consider low amount of bits to reduce feedback overhead. 
    \item Finally, different channel prediction algorithms can be considered and standardized. Furthermore, possible standardization elements are: prediction algorithm, CSI memory, and message exchanges between BS and MT.
\end{enumerate}
\section{Simulation Results}\label{results}
This section showcases the performance of the proposed scheme by means of Monte-Carlo numerical simulations. For the sake of simplicity, we  consider a single user MIMO system, composed of a single BS and a MT in a cell, equipped with $N_{t}=2$ and $N_r=1$ transmit and receive antennas, respectively. In addition, length of tapped-delay line and order of AR process is equal to $1$. We adopt the adaptive moment estimation, \textit{Adam} \cite{kingadam} optimizer to enchance the RNN, and the optimal number of hidden neurons are found to be $M_h=16$. Moreover, in order to observe only quantization effect, we assume zero-estimation error in the channel, which is estimated at MT, at time $t$. Finally, the results are scrutinized by considering mean-squared-error (MSE), which is calculated as
\begin{equation}
    \mho_{\text{mse}}={\frac{1}{T} }{\sum_{t=1}^{T}} \mid\mid{{\mathbf{H}}(t)-\widehat{\mathbf{H}}_{BS}(t)}\mid\mid^2\:
    \label{mse_comp}
\end{equation}
where, importantly, in case of conventional scheme, $\widehat{\mathbf{H}}_{BS}(t)=Q_f(\widehat{\mathbf{H}}_{MT}(t))$. 
Besides, the performance of the proposed scheme is also verified by calculating the received signal-to-noise ratio (SNR), denoted by $\Gamma_{P}$, at the MT. For this purpose, we use $\widehat{\mathbf{H}}_{BS}(t)$ to obtain a simple matched-filter (MF) precoder.
\subsection{Performance of RNN-based predictor}
To train RNN, a data-set, which is composed of a set of past CSI realizations, is extracted from consecutive $10^{4}$ data blocks, that is, $\{\widehat{\mathbf{H}}^{q}(t)|1\leq t\leq10^{4}\}$. Besides, out of $10^{4}$ data blocks, $80\%$ is used as a training, $10\%$ as a validation, and $10\%$ as a test data-set. The training process starts by initializing random weights. At training iteration, $t$, the corresponding channel matrix, i.e., $\widehat{\mathbf{H}}^{q}(t)$, is fed into the RNN along with the delayed versions, i.e., $\{\widehat{\mathbf{H}}^{q}(t-1), ..., \{\widehat{\mathbf{H}}^{q}(t-k)\}$. Afterwards, the resultant predicted channel, i.e., $\widetilde{\mathbf{H}}(t+K)$, is compared with the desired channel, i.e., ${\mathbf{\widehat{H}}^{q}}(t+K)$, and the error $\widetilde{\mathbf{H}}(t+K)-{\mathbf{\widehat{H}}^{q}}(t+K)$ is fed back to update the weights by using a dedicated training algorithm, e.g., \textit{Adam}, in our case. This iterative process ends when a predefined convergence condition is satisfied. Finally, to measure the prediction accuracy of the RNN algorithm, a test data-set is used. In this regard, MSE is considered as a performance metric, which is written as
\begin{equation}
    \eta_{\text{mse}}={\frac{1}{P} }{\sum_{p=1}^{P}} 	\mid\mid{{\widetilde{\mathbf{H}}}(p)-{\widehat{\mathbf{H}}}^{q}(p)}\mid\mid^2\:
\end{equation}
where $p$ denotes the time instant of $p^{th}$ test channel matrix and $P$ represents the total length of test data-set.
\par In our experiments, $\eta_{\text{mse}}= 6\times 10^{-3}$ and $\eta_{\text{mse}}= 7\times 10^{-3}$ are observed, for the real and imaginary part of the channel matrix, respectively. Moreover, the computational complexity of RNN-based predictor can be visualized in terms of required complex multiplication operations \cite{KFvsAI}. To this end, for single-step prediction, hidden layer will have to perform $M\cdot M_{h}$ multiplications and output layer $M_o\cdot M_{h}$. Therefore, the total required multiplications are
\begin{equation}
    \varkappa_r=(M+M_o)\cdot M_h\:.
    \label{req_mult}
\end{equation}
Additionally, the number of input neurons, $M$, are dependent on number of MIMO sub-channels and the delayed taps. Therefore, the total number of input neurons can be calculated as
\begin{equation}
 \begin{aligned}
    M&=\underbrace{N_r\cdot N_t}_\text{ $\widehat{\mathbf{H}}^q(t)$}+k\cdot\underbrace{(N_r\cdot N_t)}_\text{ $\widehat{\mathbf{H}}^q(t-k)$}+\underbrace{N_r\cdot N_t}_\text{Output feedback}\\
    &=N_r\cdot N_t(k+2)\:.
    \label{in_neuron}
    \end{aligned}
\end{equation}
On the other hand, output neurons, $M_o$, are solely dependent on MIMO sub-channels. Therefore,
\begin{equation}
    M_o=N_r\cdot N_t\:.
    \label{out_neuron}
\end{equation}
By substituting \eqref{in_neuron} and \eqref{out_neuron} into \eqref{req_mult}, the total required multiplications can be written in the simplified form as
\begin{equation}
    \varkappa_r=N_r\cdot N_t\cdot(k+3)\cdot M_h\:.
\end{equation}
\par Further, for the sake of simplicity, let us denote $\delta=N_r\cdot N_t$, which is the configuration of the MIMO system and $\varrho=k\cdot M_h$ represents the scale of RNN. Thus, the one-step prediction complexity of RNN can be written as $\mathcal{O}(\delta\varrho)$. Contrarily, training complexity is also bounded to the number of training samples, $S$, and the number of epochs, $\tau$. Thereby, the training complexity of RNN can be expressed as $\mathcal{O}(\delta\varrho S\tau)$.
\par Nevertheless, the design of an optimal channel predictor is not the objective of our work. Therefore, in the following, we evaluate the performance of conventional and proposed scheme using the metrics, $\mho_{\text{mse}}$ and $\Gamma_P$. 
\subsection{Performance of Proposed Scheme}
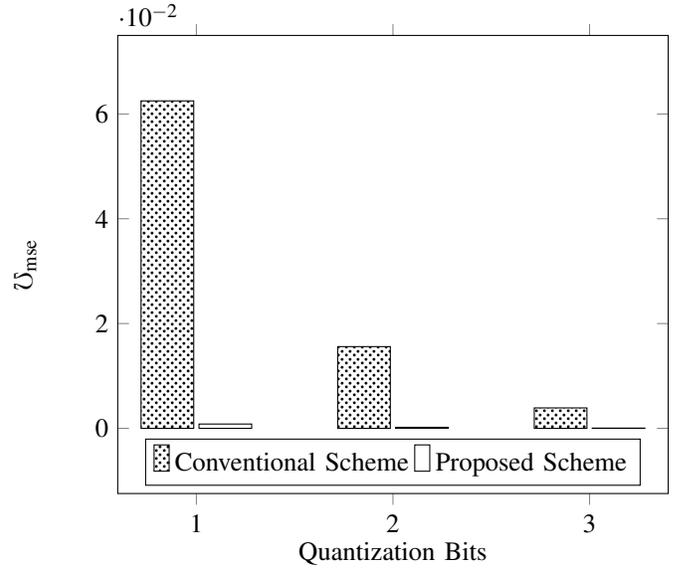
\begin{figure}
 \centering

\begin{tikzpicture}
\begin{axis}[width=8.90cm,
ybar,
enlargelimits=0.2,legend style={at={(0.95,0.12)},
cells={anchor=west}
},
legend entries={Conventional Scheme,Proposed Scheme},
bar width=0.7cm,legend columns=2,
ylabel={$\mho_{\text{mse}}$},
symbolic x coords={1,2,3},
xtick=data,
ytick={0,0.02,0.04,0.06},
xlabel={Quantization Bits},
legend image code/.code={
        \draw [#1] (0cm,-0.1cm) rectangle (0.2cm,0.25cm); },
]

\addplot [draw=black,pattern=crosshatch dots,error bars/.cd,y dir=both,y explicit,error bar style={line width=1pt}] 
coordinates
{
(1,0.0625)
(2,0.0156)
(3,0.0039)

};

;

\addplot [draw=black,error bars/.cd,y dir=both,y explicit,error bar style={line width=1pt}] 
coordinates
{
(1,0.000823)
(2,0.000196)
(3,0.0000469)
};

;

\end{axis}
\end{tikzpicture}
  \caption{{Performance analysis of conventional vs. proposed scheme in terms of MSE metric, $\mho_\text{mse}$, given in \eqref{mse_comp}}.}\label{Result_1}  
\end{figure} 
\par
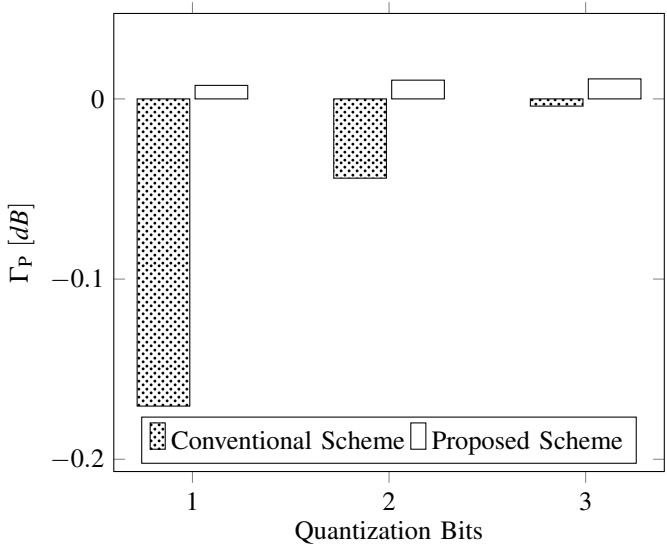
\begin{figure}
 \centering
\begin{tikzpicture}
\begin{axis}[width=8.90cm,
ybar,
enlargelimits=0.2,legend style={at={(0.95,0.12)},
cells={anchor=west}
},
legend entries={Conventional Scheme,Proposed Scheme},
bar width=0.7cm,legend columns=2,
ylabel={$\Gamma_{\text{P}}\:\:[dB]$},
symbolic x coords={1,2,3},
xtick=data,
ytick={-0.20,-0.10,0,0.05},
xlabel={Quantization Bits},
legend image code/.code={
        \draw [#1] (0cm,-0.1cm) rectangle (0.2cm,0.25cm); },
]

\addplot [draw=black,pattern=crosshatch dots,error bars/.cd,y dir=both,y explicit,error bar style={line width=1pt}] 
coordinates
{
(1,-0.1705)
(2,-0.044)

(3,-0.0040)

};

;

\addplot [draw=black,error bars/.cd,y dir=both,y explicit,error bar style={line width=1pt}] 
coordinates
{
(1,0.0075)
(2,0.0104)

(3,0.0111)

};

;

\end{axis}
\end{tikzpicture}
  \caption{{Precoding SNR, $\Gamma_P$, behaviour in proposed vs. conventional scheme.}}\label{Result_2}  
\end{figure}

 Fig.\,\ref{Result_1} reveals the trend of MSE, $\mho_\text{mse}$, for different values of quantization bits. Also, the results are portrayed for both the schemes, i.e., conventional and proposed. It can be seen that $\mho_\text{mse}$ reduces with the increase in quantization bits. However, the superior performance of the proposed scheme can be clearly observed. For instance, for one quantization bit, conventional scheme has $\mho_\text{mse}\approx 6.2\times10^{-2}$. In contrast, the proposed scheme has $\mho_\text{mse}\approx 0$; thus, the proposed scheme has reduced MSE by a huge margin. Similarly, increasing the quantization bits is reducing the MSE of the conventional scheme, and it is approaching the proposed one. On the other hand, the proposed scheme's MSE has squeezed to zero, which depicts the 100\% recovery of the estimated channel. In a nutshell, the proposed scheme not only saves the quantization bits but also reduces MSE significantly. Such reduction in MSE can greatly improve the performance of the MIMO precoder, which we investigate in Fig.\,\ref{Result_2}.  
\par Fig.\,\ref{Result_2} shows received SNR at the MT by varying number of quantization bits. It can be depicted that $\Gamma_P\approx-0.17\,dB$ and $\Gamma_P\approx7.5\times10^{-3}\,dB$ for the conventional and proposed schemes, respectively, when quantization bit is $1$. Nevertheless, in the case of $2$ and $3$ quantization bits, there is negligible change in the proposed scheme as the variation in MSE is tremendously low, which can be verified from Fig.\,\ref{Result_1}. In contrast, the conventional scheme is trying to catch up with the proposed scheme when the quantization bits are increasing; thus, remark-3, given in Section\,\ref{proposed}, holds true. By and large, the proposed scheme is outperforming by using lower number of quantization bits; thereby, reducing OTA-overhead. On the other side, the conventional scheme requires a higher number of quantization bits to achieve similar performance.       
\section{Conclusion} \label{conclusion}
This paper introduced the potential use of AI to not only reduce the overhead for CSI feedback but also to get an accurate recovery of estimated CSI at the BS. In particular, to eliminate CSI loss due to compression, a novel compression strategy is introduced. For this purpose, twin AI-enabled channel predictors are utilized at the BS and MT, which are trained on previously available compressed CSI. Simulation results showed that approximately 100\% estimated channel is recovered at the BS with the lower number of quantization bits as compared to the conventional scheme. Moreover, the achieved precoding gain verified the validity of the proposed scheme. The proposed scheme can play a significant role in scenarios where overhead and inaccurate CSI are intolerable. 
\bibliographystyle{IEEEtran}
\bibliography{mybib}
\end{document}